\documentclass[journal]{IEEEtran}

\ifCLASSINFOpdf
\else
   \usepackage[dvips]{graphicx}
\fi
\usepackage{url}

\usepackage{graphicx}
\usepackage{caption}
\usepackage{subcaption}
\usepackage{algorithm}
\usepackage{algorithmic}

\usepackage{amsmath,amssymb}
\usepackage[keeplastbox]{flushend}
\usepackage{balance}
\usepackage[inline]{enumitem}
\usepackage[all=normal,paragraphs=tight,floats=normal,mathspacing=normal,wordspacing=normal,charwidths=normal,mathdisplays=normal,leading=normal]{savetrees}

\usepackage{xcolor}
\newcommand{\new}[1]{{\leavevmode\color{black}#1}}

\begin{document}


\title{mpNet: variable depth unfolded neural network \\ for massive MIMO channel estimation}

\author{Taha Yassine, Luc Le Magoarou
\thanks{Taha Yassine and Luc Le Magoarou are with b$<>$com, Rennes, France. Taha Yassine is also with Univ. Rennes, INSA Rennes, IETR, UMR CNRS 6164, 20 av. des buttes de Coësmes, F-35708 Rennes, France. Contact addresses:  \texttt{taha.yassine@b-com.com}, \texttt{luc.lemagoarou@b-com.com}.}}

\markboth{}
{Shell \MakeLowercase{\textit{et al.}}: Bare Demo of IEEEtran.cls for IEEE Journals}
\maketitle

\begin{abstract}
Massive multiple-input multiple-output (MIMO) communication systems have a huge potential both in terms of data rate and energy efficiency, although channel estimation becomes challenging for a large number of antennas. Using a physical model allows to ease the problem by injecting a priori information based on the physics of propagation. However, such a model rests on simplifying assumptions and requires to know precisely the configuration of the system, which is unrealistic in practice.
In this paper we present mpNet, an unfolded neural network specifically designed for massive MIMO channel estimation. It is trained online in an unsupervised way. Moreover, mpNet is computationally efficient and automatically adapts its depth to the signal-to-noise ratio (SNR). The method we propose adds flexibility to physical channel models by allowing a base station (BS) to automatically correct its channel estimation algorithm based on incoming data, without the need for a separate offline training phase.
It is applied to realistic millimeter wave channels and shows great performance, achieving a channel estimation error almost as low as one would get with a perfectly calibrated system. It also allows incident detection and automatic correction, making the BS resilient and able to automatically adapt to changes in its environment.
\end{abstract}


\IEEEpeerreviewmaketitle

\section{Introduction}

\IEEEPARstart{T}{he} ever-growing need for data rate in modern communication networks led to use channels of very high dimension, which makes channel estimation difficult. For example, it has been recently proposed to use massive multiple-input multiple-output (MIMO) wireless systems \cite{Rusek2013,Larsson2014,Lu2014} with a large number of transmit and receive antennas in the millimeter-wave band \cite{Rappaport2013, Swindlehurst2014, Xiao2017}, where a large bandwidth can be exploited. In that case the channel comprises hundreds or even thousands of complex numbers, whose estimation is a very challenging signal processing problem \cite{Heath2016}.

Data processing techniques are often based on the \emph{manifold assumption}: Meaningful data (signals) lie near a low dimensional manifold, although their apparent dimension is much larger \cite{Carlsson2009,Peyre2009} \cite[Section 5.11.3]{Goodfellow2016} \cite[Section 9.3]{Elad2010}. This fact has classically been exploited in two different ways.

On the one hand, for decades if not centuries, scientists have handcrafted \emph{analytical models}. This amounts to come up with a mathematical description of the manifold, based on domain knowledge and careful observation of the phenomena of interest. This approach reaches its limits for complex phenomena that are difficult to model with a reasonable number of parameters, in which case simplifying assumptions have to be made, hindering model relevance.

On the other hand, thanks to the advent of modern computers, \emph{machine learning} (ML) techniques have emerged and led to tremendous successes in various domains \cite{Lecun2015,Goodfellow2016}. One of their main feature is to avoid any explicit mathematical description of the manifold at hand, which is taken into account via a large amount of training data sampling it. Such an approach is particularly successful in application domains for which building analytical models is difficult, since it is much more flexible. However, flexibility comes at the price of computationally heavy learning and difficulties to inject a priori knowledge on the phenomena at hand.

Recently, a promising approach meant to combine the advantages of the two aforementioned approaches has been proposed under the name of \emph{deep unfolding} (also known as deep unrolling). It amounts to unfold iterative algorithms initially based on analytical models so as to express them as neural networks that can be optimized \cite{Gregor2010,Hershey2014,Kamilov2016}. This has the advantage of adding flexibility to algorithms based on classical models, and amounts to constrain the search space of neural networks by using domain knowledge. Moreover, this leads to inference algorithms of controlled complexity (see \cite{Monga2021} and references therein for a complete survey).

Channel estimation is of paramount importance for communication systems in order to optimize the data rate/energy consumption tradeoff.
In modern systems, the possibly large number of transmit/receive antennas and subcarriers makes this task difficult. Fortunately, despite the high dimension of the channel, realistic channels are often well approximated by only a few dominant propagation paths \cite{Samimi2016} (typically less than ten). Such channels are said \emph{sparse}.
For massive MIMO channel estimation, it is thus customary to use an analytical model based on the physics of propagation in order to ease the problem. This amounts to parameterize a manifold by physical parameters such as the directions, delays and gains of the dominant propagation paths, the dimension of the manifold being equal to the number of real parameters considered in the model. Physical channel models allow injecting strong a priori knowledge based on solid principles \cite{Sayeed2002,Bajwa2010,Lemagoarou2018}, but necessarily make simplifying assumptions (e.g., the plane wave assumption \cite{Lemagoarou2019b}) and require knowing exactly the system configuration (positions of the antennas, gains, etc.). Such requirements being unrealistic in practice, the massive MIMO channel estimation task could perfectly benefit from the flexibility offered by the deep unfolding approach.

\new{Note that another way of injecting a priori knowledge is to use a bayesian channel estimator (such as the LMMSE \cite{Biguesh2006}). This has the advantage of not relying on a physical model making simplifying assumption, but requires estimating the channel distribution (which changes slowly with time). Moreover, building such a prior knowledge requires using another channel estimation method to gather estimates (often several dozens) in order to form empirical covariance matrices.}

\noindent{\bf Contributions.}
In this paper, we introduce {\sf mpNet}, an unfolded neural network specifically designed for massive MIMO channel estimation. The unfolded algorithm is matching pursuit \cite{Mallat1993}, which is a greedy computationally efficient algorithm taking advantage of the massive MIMO channel sparsity. The network takes as input the least squares (LS) channel estimates. Considering this estimate a noisy version of the channel, the objective can be seen as channel denoising. The weights of {\sf mpNet} are initialized using an imperfect physical model. The main concerns while designing the method were to make it:
\begin{itemize}
\item {\bf Unsupervised}: no need for clean channels to train {\sf mpNet}, the method uses the LS channel estimates as training data and a reconstruction cost function. It is thus trained as an autoencoder \cite{Rumelhart1986}. \new{In addition, and unlike the minimum mean squared error (MMSE) estimator, {\sf mpNet} does not require knowledge of the channel correlation matrix.}
\item {\bf Online}: the traditional offline training phase is replaced by an intelligent initialization using an imperfect physical model. {\sf mpNet} is then trained incrementally using online gradient descent \cite{Bottou1998}.
\item {\bf SNR adaptive}: no need for several networks trained at different signal-to-noise ratio (SNR) levels, {\sf mpNet} automatically adapts its depth to incoming data.
\item {\bf Computationally efficient}: backpropagation through {\sf mpNet} is cheaper than classical channel estimation using a greedy sparse recovery algorithm (which corresponds to the forward pass of {\sf mpNet}).
\end{itemize}
To the best of the authors' knowledge, the proposed method is the only one meeting all these requirements. Such a method is particularly suited to imperfectly known or non-calibrated systems. Indeed, starting from an imperfect physical channel model, our method allows a base station (BS) to automatically correct its channel estimation algorithm based on incoming data.

\noindent{\bf Related work.} ML holds promise for wireless communications (see \cite{Oshea2017,Wang2017,Qin2019} for exhaustive surveys). More specifically, since the physics of propagation provides pretty accurate analytical models, model-driven ML approaches \cite{He2019} seem particularly suited. The method we propose can be seen as an instance of the model-driven approach.

In the context of massive MIMO channel estimation, it has recently been proposed to use adaptive data representations using dictionary learning techniques \cite{Ding2018}. However, classical dictionary learning employing algorithms such as K-SVD \cite{Aharon2006}, as proposed in \cite{Ding2018}, is very computationally heavy, and thus not suited to online learning, as opposed to the {\sf mpNet} approach.

Deep unfolding has also been considered by communications researchers (see \cite{Balatsoukas2019} for a survey). It has mainly been used for symbol detection, unfolding projected gradient descent \cite{Samuel2017} or approximate message passing (AMP) \cite{He2018a} algorithms. 

Regarding MIMO channel estimation, it has been proposed in \cite{He2018} to unfold a sparse recovery algorithm named denoising-based AMP (DAMP) \cite{Metzler2016}. However, the method is directly adapted from image processing \cite{Metzler2017} and does not make use of a physical channel model as initialization as we propose here. A recent work also proposes to use deep unfolding of the DAMP algorithm for MIMO channel estimation \cite{Wei2019}, using a physical model to optimize the shrinkage functions used by DAMP. However, these previously proposed methods all require collecting a database of clean channels and an offline training phase, due to their intrinsic supervised nature. This may hinder their practical applicability. Moreover, the proposed unfolded neural networks are much more computationally complex than {\sf mpNet}, and do not comprise an automatic way to adapt to the SNR.

Finally, deep learning has also been applied to channel estimation in an orthogonal frequency-division multiplexing (OFDM) context, which is mathematically very close to the one studied here. In \cite{Gao2018}, a neural network is used as a post-treatment of the LS channel estimates in order to denoise them, and included in a joint channel estimation and detection framework. In \cite{Soltani2019}, the noisy time-frequency response of the channel is viewed as an image  and is denoised using classical denoising neural networks. Once again, these approaches are supervised and are of high complexity compared to classical methods \cite{Vanlier2020}.

\new{In \cite{Balevi2020}, a low complexity neural network based on the deep image prior network is used as a pre-treatment of the LS channel estimates and can be seen as a denoiser. This approach shares similarities with ours as the training is done online and in an unsupervised manner, but differs mainly in three aspects:
\begin{enumerate}
	\item Our proposed method acts on LS estimates (as a post-treatment) whereas the method of \cite{Balevi2020} acts before correlation with the pilot sequences (as a pre-treatment of LS estimation).

	\item The way the neural network is handled is totally different for the two methods. Indeed, we consider a network taking as input a noisy channel, and weights parameterizing a channel denoising operation that are shared by all users in the network and stay constant over large periods of time (as long as the environment does not change significantly). Once the network is properly trained, no gradient descent is required anymore when encountering a new channel realization. On the other hand,the method of \cite{Balevi2020} considers a neural network taking a constant vector as input and optimizes its weights \emph{for every channel realization}. In that case, gradient descent is always required to solve the above problem as long as the method is used (since learnt parameters are only for a single realization).

	\item Our method uses a priori knowledge about physical parameters in order to structure the proposed neural network (using deep unfolding) and to initialize it in a clever way. A continuous flow of data comes then to further optimize the neural network in order to mitigate any error induced by small uncertainties on the physical parameters. By contrast, the method in \cite{Balevi2020} considers a general approach to signal denoising, \emph{without any physical assumption about the problem at hand}. Both methods can be seen as projections of noisy channels onto low dimensional manifolds in order to denoise (channels are preserved while noise is suppressed by the projection). However, the method of \cite{Balevi2020} assumes very little about the specific form of the manifold, that is encoded by the structure of the network in order to introduce correlations between neighboring channel coefficients (as originally proposed for image data in \cite{Heckel2018}). On the opposite, our method uses the physics of wave propagation to constrain the manifold onto which corrupted channels are projected.
\end{enumerate}}


\noindent{\bf Organization.} The remaining of the paper is organized as follows. First, section~\ref{sec:problem_formulation} introduces the problem at hand and describes the physical model on which {\sf mpNet} is based. Then, the motivations behind the proposed solution is presented in section~\ref{sec:impact}. Section~\ref{sec:prop_approach} introduces in details {\sf mpNet}: the deep unfolding based strategy we propose for MIMO channel estimation. In section~\ref{sec:experiments}, different experiments are conducted in order to assess and validate the potential of our approach. Finally, section~\ref{sec:conclusion} discusses the contributions and concludes the paper.

\section{Problem formulation}
\noindent\new{{\bf Notations.} Matrices and vectors are denoted by bold upper-case and lower-case letters: $\mathbf{A}$ and $\mathbf{a}$ (except 3D ``\emph{spatial}’’ vectors that are denoted $\overrightarrow{a}$). 
A matrix transpose 
and transconjugate is denoted by $\mathbf{A}^T$ 
and $\mathbf{A}^H$ respectively.
The identity matrix is denoted by $\mathbf{Id}$.
$\mathcal{CN}(\boldsymbol\mu,\boldsymbol{\Sigma})$ denotes the standard complex gaussian distribution with mean $\boldsymbol\mu$ and covariance $\boldsymbol{\Sigma}$. $\mathbb{E}(\cdot)$ denotes the expectation.
$\Vert\cdot\Vert_2$ and $\vert\cdot\vert$ denote the L2-norm and the complex modulus respectively.}

\label{sec:problem_formulation}
\subsection{System settings}
We consider in this paper a massive MIMO system, also known as multi-user MIMO (MU-MIMO) system \cite{Rusek2013,Larsson2014,Lu2014}, in which a BS equipped with $N$ antennas communicates with $M$ single antenna users ($M<N$). Let us consider for ease of presentation a transmission on a single subcarrier, even though everything presented in the paper can obviously be generalized to the multicarrier case. The system operates in time division duplex (TDD) mode, so that channel reciprocity holds and the channel is estimated in the uplink: each user sends a pilot sequence $\mathbf{p}_m \in \mathbb{C}^{T}$ of duration $T$ (orthogonal to the sequences of the other users, $\mathbf{p}_m^H\mathbf{p}_l = \delta_{ml}$ where $\delta_{ml}$ is the Kronecker delta) for the BS to estimate the channel.\new{For simplicity, the transmit power is set to $1$ throughout the paper, and the signal to noise ratio (SNR) is determined only by the channel and noise powers.} The received signal is thus expressed 
\begin{equation}
\mathbf{R}=\sum_{m=1}^M\mathbf{h}_m\mathbf{p}_m^H +\mathbf{N},
\end{equation} 
where $\mathbf{R}\in \mathbb{C}^{N \times T}$, $\mathbf{h}_m$ is the channel corresponding to the $m$th user and $\mathbf{N}\in \mathbb{C}^{N \times T}$ is Gaussian noise.
After correlating the received signal with the pilot sequences, and assuming no pilot contamination from adjacent cells for simplicity, the BS gets noisy measurements of the channels of all users, taking the form 
\begin{equation}
\mathbf{x}_l \triangleq \mathbf{R}\mathbf{p}_l =\underbrace{\sum\nolimits_{m=1}^M\mathbf{h}_m\mathbf{p}_m^H\mathbf{p}_l}_{\mathbf{h}_l} + \underbrace{\mathbf{N}\mathbf{p}_l}_{\mathbf{n}_l},
\end{equation}
for $l=1,\dots,M$, with $\mathbf{n}_l \sim \mathcal{CN}(0,\sigma^2\mathbf{Id}),\, \forall l$. 
In order to simplify notations in the remaining parts of the paper, we drop the user index and denote such measurements with the canonical expression 
\begin{equation}
\mathbf{x} = \mathbf{h} + \mathbf{n},
\label{eq:observations}
\end{equation}
where $\mathbf{h}$ is the channel of the considered user and $\mathbf{n}$ is noise, with $\mathbf{n} \sim \mathcal{CN}(0,\sigma^2\mathbf{Id})$. Such a dropping of the user index does not harm the description of our approach, since it treats the channels of all users indifferently. Note that $\mathbf{x}$ is already an unbiased estimator of the channel, obtained by solving a LS estimation problem, so that we call it the LS estimator in the sequel. Its performance can be assessed by the input SNR
$$
\text{SNR}_{\text{in}}  \triangleq \frac{\left\Vert \mathbf{h} \right\Vert_2^2}{N\sigma^2}.
$$ 
However, one can get better channel estimates using a physical model which allows to denoise the LS estimate, as is explained in the next subsection.

\subsection{Physical model}
\new{Let us denote $\{g_1,\dots,g_N\}$ the complex gains of the BS's antennas, $\{\overrightarrow{a_1},\dots,\overrightarrow{a_N}\}$ their positions with respect to the centroid of the antenna array and $\lambda$ the considered wavelength.} Then, under the plane wave assumption and assuming omnidirectional antennas (isotropic radiation patterns), the channel resulting from a single propagation path with direction of arrival (DoA) $\overrightarrow{u}$ is proportional to the \emph{steering vector}
\new{
$$
\mathbf{e}(\overrightarrow{u}) \triangleq \frac{1}{\sqrt{\sum^N_{i=1}\vert g_i\vert^2}}(
g_1\mathrm{e}^{-\mathrm{j}\frac{2\pi}{\lambda}\overrightarrow{a_1}.\overrightarrow{u}}, \dots, g_N\mathrm{e}^{-\mathrm{j}\frac{2\pi}{\lambda}\overrightarrow{a_N}.\overrightarrow{u}})^T 
$$}
which reads $\mathbf{h} = \beta \mathbf{e}(\overrightarrow{u}),$
with $\beta \in \mathbb{C}$. In that case, a sensible estimation strategy \cite{Sayeed2002,Bajwa2010,Lemagoarou2018} is to build a dictionary of steering vectors corresponding to $A$ potential DoAs: $\mathbf{E} \triangleq \begin{pmatrix}
\mathbf{e}(\overrightarrow{u_1}),\dots,\mathbf{e}(\overrightarrow{u_A})
\end{pmatrix}
$ and to compute a channel estimate with the procedure
\begin{subequations}\label{eq:estim_strat}
\begin{align}
&\overrightarrow{v}=\text{argmax}_{\overrightarrow{u_i}} \,\,\, |\mathbf{e}(\overrightarrow{u_i})^H\mathbf{x}|,\label{eq:estim_strat1}\\
&\hat{\mathbf{h}} = \mathbf{e}(\overrightarrow{v})\mathbf{e}(\overrightarrow{v})^H\mathbf{x}.\label{eq:estim_strat2}
\end{align} 
\end{subequations}
The first step, \eqref{eq:estim_strat1}, of this procedure amounts to find the column of the dictionary the most correlated with the observation $\mathbf{x}$ (LS channel estimate) to estimate the DoA $\overrightarrow{v}$ while the second step, \eqref{eq:estim_strat2}, amounts to project the observation on the corresponding steering vector $\mathbf{e}(\overrightarrow{v})$. The SNR at the output of this procedure reads
$$\text{SNR}_\text{out} \triangleq \frac{\Vert \mathbf{h} \Vert_2^2}{\mathbb{E}\big[\Vert\mathbf{h}-\hat{\mathbf{h}}\Vert_2^2\big]},$$
and we have at best $\text{SNR}_\text{out} =N\text{SNR}_\text{in}$ (neglecting the discretization error), if the selected steering vector is collinear to the actual channel. This is a direct consequence of the Cauchy-Schwarz inequality, and is intuitively explained by the fact that from the $N$ complex dimensions of $\mathbf{x}$ corrupted by noise, only one is kept when projecting on the best steering vector, so that the effective noise variance is divided by $N$. The potential gain of using such a physical model can be huge, especially for massive MIMO systems in which the number of antennas $N$ is large.
 
Moreover, this strategy can be generalized to estimate sparse multipath channels of the form
\begin{equation}
\mathbf{h} = \sum\nolimits_{p=1}^P\beta_p \mathbf{e}(\overrightarrow{u_p}),
\label{eq:multipath_channel}
\end{equation}
\new{where $P$ is the number of paths}, by iterating the procedure \eqref{eq:estim_strat} until some predefined stopping criterion is met. This leads to greedy sparse recovery algorithms such as matching pursuit (MP) \cite{Mallat1993} or orthogonal matching pursuit (OMP) \cite{Tropp2010}. Since the method proposed in this paper is based on the unfolding of the matching pursuit algorithm, a high level overview of it applied to channel estimation, with dictionary $\mathbf{E}$ and input $\mathbf{x}$ is given in algorithm~\ref{alg:mp}.

\begin{algorithm}[htb]
\caption{Matching pursuit \cite{Mallat1993} (high level overview)}
\begin{algorithmic}[1] 
\REQUIRE Dictionary $\mathbf{E}$, input $\mathbf{x}$ (noisy channel)
\STATE $\mathbf{r} \leftarrow \mathbf{x}$
\WHILE{Stopping criterion not met}
\STATE Find the most correlated atom: $s\leftarrow\underset{i}{\text{argmax}} \,\,\, |\mathbf{e}_i^H\mathbf{r}|$
\STATE Update the residual: $\mathbf{r} \leftarrow \mathbf{r} - \mathbf{e}_s\mathbf{e}_s^H\mathbf{r}$
\ENDWHILE
\ENSURE $\hat{\mathbf{h}} \leftarrow \mathbf{x} - \mathbf{r}$ (denoised channel)
\end{algorithmic}
\label{alg:mp}
\end{algorithm}

\section{Motivation: imperfect models}
\label{sec:impact}
Basing an estimation strategy on a physical model, as suggested in the previous section, requires knowing precisely the physical parameters of the system (in particular the positions and gains of the antennas) in order to build an appropriate dictionary. Then, even in the case of perfect system knowledge, some simplifying hypotheses (such as the plane wave assumption considered in the previous section) have to be made in order to keep the model mathematically tractable. Consequently, every model, regardless of its sophistication, is necessarily imperfect. Such a situation is well-known and summarized by the aphorism \emph{``All models are wrong''} \cite{Box1976}.

\begin{figure}[htbp]
    \centering
    \includegraphics[width=0.9\columnwidth]{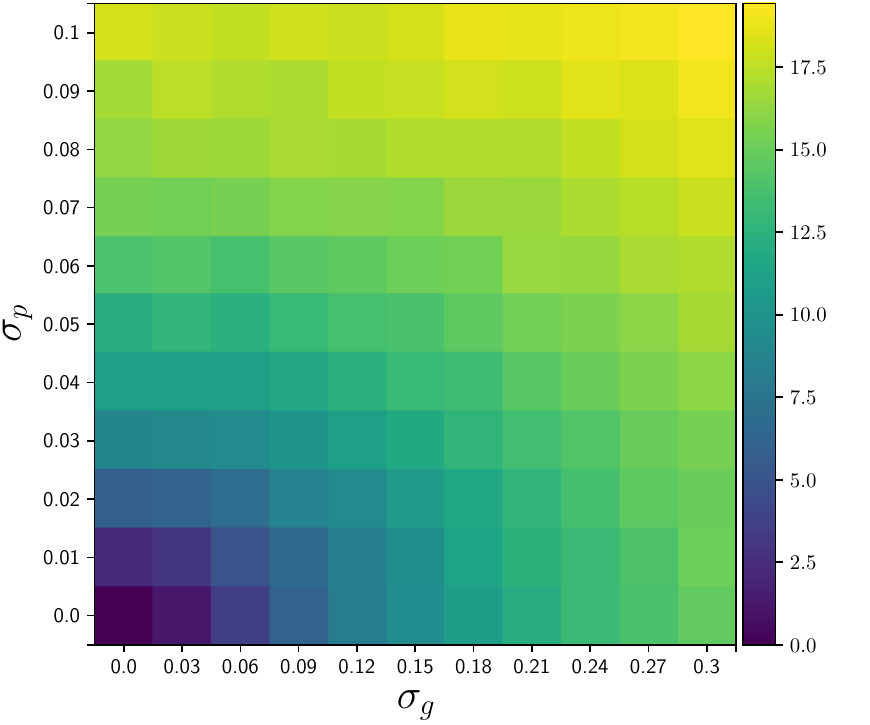}    
    \caption{SNR loss in decibels (dB) due to imperfect knowledge of the system.}
    \label{fig:imperfect_model} 
\end{figure}

In the context of MIMO channel estimation, what is the impact of an imperfect knowledge of the physical parameters and/or of the invalidity of some hypotheses? In order to address this question, let us perform a simple experiment. 
Consider an antenna array of $N=64$ antennas at the BS, whose known \emph{nominal} configuration is a uniform linear array (ULA) of unit gain antennas separated by half-wavelengths and aligned with the $x$-axis. This nominal configuration corresponds to gains and positions $\{\tilde{g}_i,\tilde{\overrightarrow{a_i}}\}_{i=1}^N$. Now, suppose the knowledge of the system configuration is imperfect, meaning that the unknown \emph{true} configuration of the system is given by the gains and positions $\{g_i,\overrightarrow{a_i}\}_{i=1}^N$, with 
\begin{align}
\begin{split}
&g_i = \tilde{g}_i + n_{g,i}, \, n_{g,i} \sim \mathcal{CN}(0,\sigma_g^2),\\
&\overrightarrow{a_i} = \tilde{\overrightarrow{a_i}} + \lambda\mathbf{n}_{p,i}, \, \mathbf{n}_{p,i} = {\small\begin{pmatrix} e_{p,i}, &0, & 0 \end{pmatrix}^T}, e_{p,i} \sim \mathcal{N}(0,\sigma_p^2).
\end{split}
\label{eq:imperfection}
\end{align}
This way, $\sigma_g$ (resp. $\sigma_p$) quantifies the uncertainty about the antenna gains (resp. spacings). Moreover, let
\new{
$$
\tilde{\mathbf{e}}(\overrightarrow{u}) \triangleq \frac{1}{\sqrt{\sum^N_{i=1}\vert \tilde{g}_i\vert^2}} (
\tilde{g}_1\mathrm{e}^{-\mathrm{j}\frac{2\pi}{\lambda}\tilde{\overrightarrow{a_1}}.\overrightarrow{u}},\dots,
\tilde{g}_N\mathrm{e}^{-\mathrm{j}\frac{2\pi}{\lambda}\tilde{\overrightarrow{a_N}}.\overrightarrow{u}}
)^T
$$
}
be the nominal steering vector and $\tilde{\mathbf{E}} \triangleq \begin{pmatrix}
\tilde{\mathbf{e}}(\overrightarrow{u_1}),\dots,\tilde{\mathbf{e}}(\overrightarrow{u_A})
\end{pmatrix}$ be a dictionary of nominal steering vectors. The experiment consists in comparing the estimation strategy of \eqref{eq:estim_strat} using the true (perfect but unknown) dictionary $\mathbf{E}$ with the exact same strategy using the nominal (imperfect but known) dictionary $\tilde{\mathbf{E}}$. To do so, we generate measurements according to \eqref{eq:observations} with channels of the form $\mathbf{h} = \mathbf{e}(\overrightarrow{u})$ where $\overrightarrow{u}$ corresponds to azimuth angles chosen uniformly at random, and $\text{SNR}_{\text{in}}$ is set to $10\,\text{dB}$. Then, the dictionaries $\mathbf{E}$ and $\tilde{\mathbf{E}}$ are built by choosing $A=32N$ directions corresponding to evenly spaced azimuth angles. Let $\hat{\mathbf{h}}_{\mathbf{E}}$ be the estimate obtained using $\mathbf{E}$ in \eqref{eq:estim_strat}, and $\hat{\mathbf{h}}_{\tilde{\mathbf{E}}}$ the estimate obtained using $\tilde{\mathbf{E}}$. The SNR loss (performance decrease) caused by using the imperfect but known dictionary $\tilde{\mathbf{E}}$ instead of the perfect but unknown dictionary $\mathbf{E}$ is measured by the quantity 
$$
\frac{\Vert \hat{\mathbf{h}}_{\tilde{\mathbf{E}}} - \mathbf{h} \Vert_2^2}{\Vert \hat{\mathbf{h}}_\mathbf{E} - \mathbf{h} \Vert_2^2}.
$$ 
Results in terms of SNR loss, in average over $10$ antenna array realizations and $1000$ channel realizations per antenna array realization are shown on figure~\ref{fig:imperfect_model}. From the figure, it is obvious that even a relatively small uncertainty about the system configuration can cause a great SNR loss. For example, an uncertainty of $0.03\lambda$ on the antenna spacings and of $0.09$ on the antenna gains leads to an SNR loss of more than $10\,\text{dB}$, which means that the mean squared error undergoes a more than tenfold increase. 

This short experiment simply highlights the fact that using imperfect models can severely harm MIMO channel estimation performance. Note that we took here as an example of imperfection the uncertainty about antenna positions and gains, but many other sources of imperfection will impact real world practical systems. The main contribution of this paper is to propose a method that takes into account and corrects to some extent imperfect physical models using ML, and more specifically deep unfolding.

\section{Proposed approach: mpNet}
\label{sec:prop_approach}
Let us now propose a strategy based on deep unfolding allowing to correct a channel estimation algorithm based on an imperfect physical model incrementally, via online learning.

\subsection{Unfolding matching pursuit}
\label{ssec:unfolding_MP}
 
The estimation strategy summarized in algorithm~\ref{alg:mp} can be unfolded as a neural network taking the observation $\mathbf{x}$ as input and outputting a channel estimate $\hat{\mathbf{h}}$. Indeed, the first step in the while loop amounts to perform a linear transformation (multiplying the input by the matrix $\mathbf{E}^H$) followed by a nonlinear one (finding the inner product of maximum amplitude and setting all the others to zero) and the second step corresponds to a linear transformation (multiplying by the matrix $\mathbf{E}$). Such a strategy is parameterized by the dictionary of steering vectors $\mathbf{E}$. In the case where the optimal dictionary $\mathbf{E}$ is unknown (or imperfectly known), we propose to learn the dictionary matrix used in \eqref{eq:estim_strat} directly on data via backpropagation \cite{Rumelhart1986}.

\begin{figure}[htbp]
\includegraphics[width=\columnwidth]{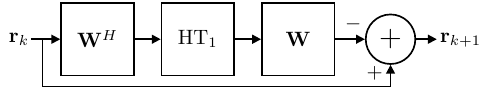}
\caption{One layer of {\sf mpNet}.}
\label{fig:mpnetK}
\end{figure}

\noindent{\bf Neural network structure.} This is done by considering the dictionary matrix as weights of the neural network we introduce, called {\sf mpNet}, whose forward pass is given in algorithm~\ref{alg:forwardmpnet}. The notation $\text{HT}_1$ refers to the hard thresholding operator which keeps only the entry of greatest modulus of its input and sets all the others to zero. The parameters of this neural network are the weights $\mathbf{W} \in \mathbb{C}^{N \times A}$, where $A$ is an hyperparameter denoting the number of considered atoms in the dictionary. Note that complex weights and inputs are handled classically by stacking the real and imaginary parts for vectors and using the real representation for matrices. The forward pass of {\sf mpNet} can be seen as a sequence of $K$ iterations whose schematic description is shown on figure~\ref{fig:mpnetK}. The stopping criterion determining the number $K$ of replications of the aforementioned structure, which corresponds to the number of estimated paths, is studied in section~\ref{ssec:stopping_criterion}, with the objective to make the depth of {\sf mpNet} adaptive to the SNR.

\begin{algorithm}[htb]
\caption{Forward pass of {\sf mpNet}}
\begin{algorithmic}[1] 
\REQUIRE Weight matrix $\mathbf{W}\in \mathbb{C}^{N\times A}$, input $\mathbf{x}$ 
\STATE $\mathbf{r} \leftarrow \mathbf{x}$
\WHILE{Stopping criterion not met}
\STATE $\mathbf{r} \leftarrow \mathbf{r} - \mathbf{W} \, \text{HT}_1(\mathbf{W}^H\mathbf{r})$
\ENDWHILE
\ENSURE $\texttt{FW}(\mathbf{W},\mathbf{x}) \leftarrow \mathbf{x} - \mathbf{r}$
\end{algorithmic}
\label{alg:forwardmpnet}
\end{algorithm}

\subsection{Training mpNet}
\label{ssec:training_mpnet}
The method we propose to jointly estimate channels while simultaneously correcting an imperfect physical model is summarized in algorithm~\ref{alg:learningmpnet}. Note that {\sf mpNet} is fed with normalized inputs of the form $\frac{\mathbf{x}}{\Vert \mathbf{x}\Vert_2}$, since we noticed it improved its performance. \new{The output of the network is then multiplied by $\Vert \mathbf{x}\Vert_2$ in order to compensate for this initial normalization to get channel estimates of correct norm (see line 5 of algorithm~\ref{alg:learningmpnet}).} The training strategy amounts to initialize the weights of {\sf mpNet} with a dictionary of nominal steering vectors $\tilde{\mathbf{E}}$ and then to perform a minibatch gradient descent \cite{Bottou2010} on the weights $\mathbf{W}$ to minimize the risk 
\begin{equation}
R \triangleq \mathbb{E} \left[\frac{1}{2} \frac{\Vert \mathbf{x} - \hat{\mathbf{h}} \Vert_2^2}{\Vert \mathbf{x}  \Vert_2^2}\right].
\end{equation}
Denoting $\mathcal{B}$ the current minibatch of size $B$, the expectation involved in the risk is approximated by computing an average over the minibatch observations, leading to the cost function
\begin{equation}
C \triangleq \frac{1}{2B}\sum_{\mathbf{x} \in \mathcal{B}} \frac{\Vert \mathbf{x} - \hat{\mathbf{h}} \Vert_2^2}{\Vert \mathbf{x}  \Vert_2^2}.
\label{eq:cost}
\end{equation}
Note that this cost function evolves with time, since all minibatches are made of different observations, thus allowing real-time adaptation of {\sf mpNet} to changes in the channel distribution.
The network is trained to minimize the average discrepancy between its inputs and outputs, exactly as a classical autoencoder. It operates \emph{online}, on streaming observations $\mathbf{x}_t,\, t=1,\dots,\infty$ of the form \eqref{eq:observations} acquired over time (coming from all users indifferently). Note that, as opposed to the classical unfolding strategies \cite{Gregor2010,Hershey2014,Kamilov2016}, the proposed method is totally \emph{unsupervised}, meaning that it requires only noisy channel observations and no database of clean channels to run. Note that in all the experiments performed in this paper, we use minibatches of $B=200$ observations and the Adam optimization algorithm \cite{Kingma2014}. By abuse of notation, we denote $\texttt{Adam}(\mathbf{W},\frac{\partial C}{\partial \mathbf{W}},\alpha)$ the update of the weights $\mathbf{W}$ by the Adam algorithm on cost function $C$ using the learning rate $\alpha$.
\begin{algorithm}[htb]
\caption{Online training of {\sf mpNet}}
\begin{algorithmic}[1] 
\REQUIRE Nominal dictionary $\tilde{\mathbf{E}}\in \mathbb{C}^{N\times A}$,  minibatch size $B$, learning rate $\alpha$
\STATE Initialize the weights: $\mathbf{W}\leftarrow \tilde{\mathbf{E}}$
\STATE Initialize the cost function: $C\leftarrow 0$
\FOR{$t=1,\dots,\infty$}
\STATE Get observation $\mathbf{x}_t$ following \eqref{eq:observations}
\STATE Estimate the channel (forward pass):\newline $\hat{\mathbf{h}}_t \leftarrow \Vert \mathbf{x}_t \Vert_2 \, \texttt{FW}\left(\mathbf{W},\frac{\mathbf{x}_t}{\Vert \mathbf{x}_t \Vert_2}\right)$
\STATE Increment the cost function:\newline $C \leftarrow C + \frac{1}{2B\Vert \mathbf{x}_t \Vert_2^2}\Vert \mathbf{x}_t - \hat{\mathbf{h}}_t \Vert_2^2$
\IF{$t \mod B = 0$}
\STATE Update the weights (backward pass):\newline $\mathbf{W} \leftarrow \texttt{Adam}(\mathbf{W},\frac{\partial C}{\partial \mathbf{W}},\alpha)$
\STATE Reset the cost function: $C\leftarrow 0$
\ENDIF
\ENDFOR
\end{algorithmic}
\label{alg:learningmpnet}
\end{algorithm}

\begin{figure*}[t]
	\captionsetup{font=footnotesize}
	\begin{subfigure}[b]{0.333\textwidth}
	\includegraphics[width=\columnwidth]{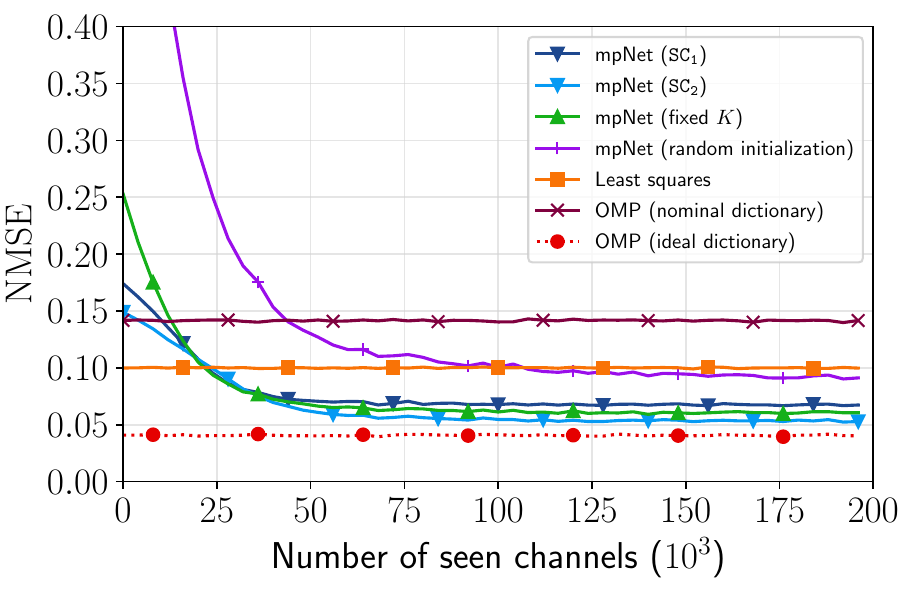}
	\caption{$\textrm{{SNR}}_{{\textrm{{in}}}}={10}\,\textrm{{dB}},\, \sigma_p={0.1},\, \sigma_g={0.3},\, K={8}$}\label{fig:multi_a}
	\end{subfigure}
	\begin{subfigure}[b]{0.333\textwidth}
	\includegraphics[width=\columnwidth]{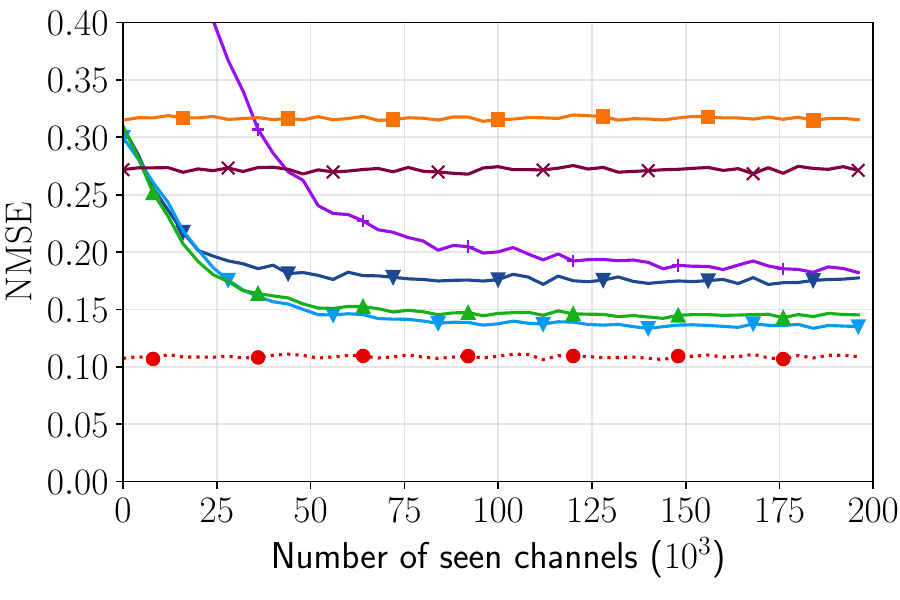}
	\caption{$\textrm{{SNR}}_{{\textrm{{in}}}}={5}\,\textrm{{dB}},\, \sigma_p={0.1},\, \sigma_g={0.3},\, K={6}$}\label{fig:multi_b}
	\end{subfigure}
	\begin{subfigure}[b]{0.333\textwidth}
	\includegraphics[width=\columnwidth]{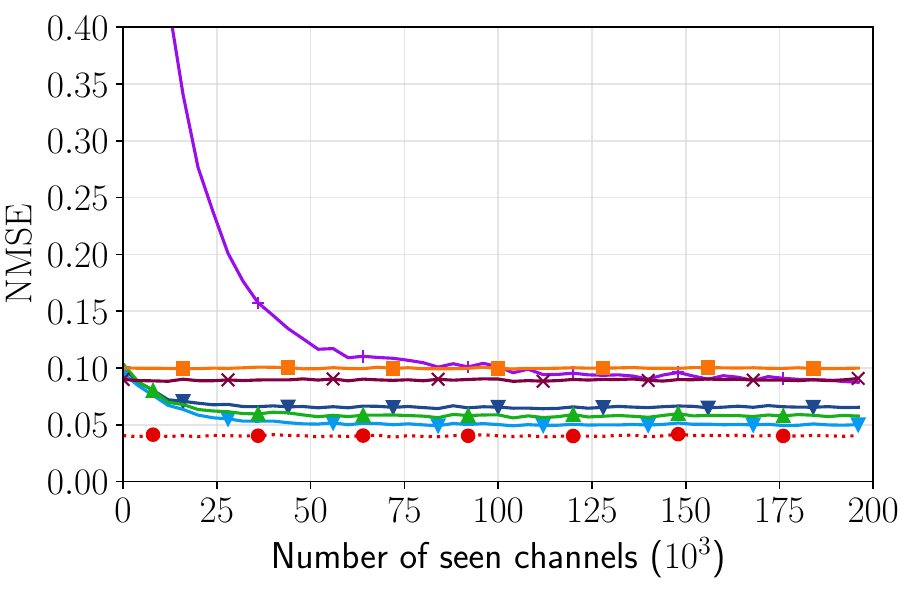}
	\caption{$\textrm{{SNR}}_{{\textrm{{in}}}}={10}\,\textrm{{dB}},\, \sigma_p={0.05},\, \sigma_g={0.15},\, K={8}$}\label{fig:multi_c}
	\end{subfigure}
	\caption{Channel estimation performance on synthetic realistic channels for various SNRs and model imperfections.}
	\label{fig:multi}
\end{figure*}

\noindent{\bf Computational complexity.} Let us denote $K$ the number of times  line~$3$ of algorithm~\ref{alg:forwardmpnet} is executed, which corresponds to the number of estimated channel paths. This number depends on the chosen stopping criterion, which is studied in section~\ref{ssec:stopping_criterion}. The forward pass of {\sf mpNet} costs $\mathcal{O}(KNA)$ arithmetic operations, its complexity being dominated by the multiplication of the input by the matrix $\mathbf{W}^H$ (first block of figure~\ref{fig:mpnetK}).
The backpropagation step costs only $\mathcal{O}(KN)$ arithmetic operations ($A$ times less), since the error flows through only one columns of the weight matrix $\mathbf{W}$ at each step, due to the hard thresholding operation done during the forward pass. This short complexity analysis means that jointly  estimating the channel (forward pass) and learning the model (backward pass)  is done at a cost that is overall the same order as the one of simply estimating the channel with a greedy algorithm (MP or OMP), without adapting the model at all to data (which corresponds to computing only the forward pass). This very light computational cost makes the method particularly well suited to online learning, as opposed to previously proposed channel estimation strategies based on deep learning. Indeed, \cite{He2018} and \cite{Wei2019}, for example, are both based on modified versions of AMP where certain steps (e.g., shrinkage function) were replaced by trainable neural networks. If AMP (\cite{Donoho2009}) were to be used in similar settings to ours, its computational complexity would be $\mathcal{O}(K'NA)$ which is of the same order as that of MP. Knowing that AMP needs more iterations $K'$ to converge than MP (i.e., $K'>K$), this means that it is indeed more computationally heavy. Moreover, modified versions of AMP could be even slower if their modified parts are particularly computationally demanding. In fact, the denoiser neural network used in \cite{He2018} as a replacement for the shrinkage function has alone way more trainable parameters than our network (20 convolutional layers 18 of which use 64 different $3\times3\times64$ filters). \new{Finally, the fully-connected denoiser neural network used in \cite{Vanlier2020} as a post-treatment of the LS channel estimate has a reasonably low computational complexity when considering only the forward pass. However, when taking account of the offline training overhead, the complexity becomes higher and unsuitable for online channel estimation in a continuously varying environment.}

\subsection{Choosing the stopping criterion}
\label{ssec:stopping_criterion}

{\sf mpNet} is the unfolded version of the matching pursuit algorithm, and its number $K$ of layers corresponds to the number of iterations of the said algorithm. It also represents the number of estimated channel paths. In fact, $K$ is nothing else but a hyperparameter that needs to be optimized. But how can we determine this number appropriately?
In the preliminary version of this study \cite{Lemagoarou2020}, this was done by testing different values of $K$ and choosing the one yielding the best results in terms of relative error in average over channel observations, by cross-validation. In that case, the number of estimated paths was the same for every channel observation. For practical systems, this strategy is suboptimal. Indeed, the number of estimated paths should ideally depend on the SNR: the higher the SNR the more paths can be estimated reliably. Since the SNR depends on the distance separating the users and the BS on one hand, and the path loss and gains of the different propagation paths on the other hand, the depth of {\sf mpNet} should be allowed to vary in order to estimate a number of paths adapted to each channel observation. 

%

In order to do so, let us take advantage of the adaptive stopping criteria proposed for greedy sparse recovery algorithms. In \cite{Cai2011,Wu2012}, the authors show that OMP with the stopping criterion
\begin{equation}\label{eq:sc_old}
\mathtt{SC_1}: \,\|\mathbf{r}\|_2^2 \leq \tilde{\sigma}^2 (N+2\sqrt{N\log N}),
\end{equation}
with $\tilde{\sigma}^2\triangleq\frac{\sigma^2}{\|\mathbf{x}\|_2^2}$, is optimal in a support recovery sense. Moreover, for a small number of  iterations and an incoherent enough dictionary, MP and OMP give very close results. Hence, we propose to use $\mathtt{SC_1}$ as stopping criterion for {\sf mpNet}. Implementing this stopping criterion requires knowing the noise variance $\sigma^2$ or at least having an estimate $\hat{\sigma}^2$. Fortunately, the noise variance can be estimated quite reliably in MIMO systems \cite{Das2012}. In the sequel, we assume that a perfect noise variance estimate is available ($\hat{\sigma}^2 = \sigma^2$).

Note that practical channels $\mathbf{h}$ and learned weights $\mathbf{W}$ do not follow exactly the generative model used in \cite{Cai2011,Wu2012} to derive the optimal stopping criterion, so that $\mathtt{SC_1}$ may lose its optimality in the case studied here. For this reason, we also propose to use the simpler and more intuitive stopping criterion
\begin{equation}\label{eq:sc_new}
\mathtt{SC_2}: \,\|\mathbf{r}\|_2^2 \leq \tilde{\sigma}^2 N.
\end{equation}

From a neural network perspective, using the stopping criteria $\mathtt{SC_1}$ and $\mathtt{SC_2}$ means training a neural network whose depth is adaptive and dynamically adjusted during learning and inference. The structure of figure~\ref{fig:mpnetK} is indeed replicated until the chosen stopping criterion is met.
The two stopping criteria $\mathtt{SC_1}$ and $\mathtt{SC_2}$ are empirically compared on realistic synthetic channels in section~\ref{sec:experiments}.



\begin{figure}[htp]
\centering
\subfloat[$\mathtt{SC_1}$]{%
	\includegraphics[clip,width=0.85\columnwidth]{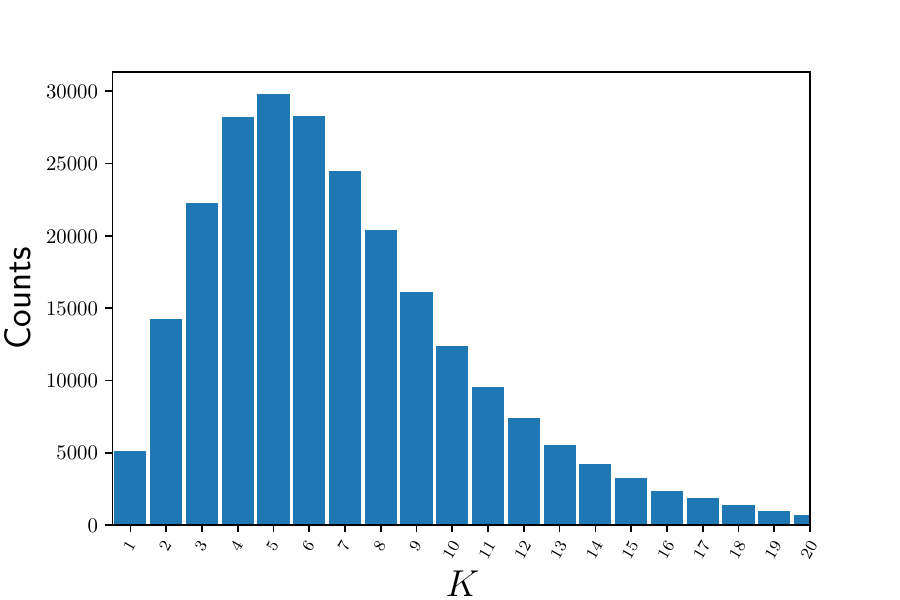}%
	\label{fig:old_K_histo}
}

\subfloat[$\mathtt{SC_2}$]{%
	\includegraphics[clip,width=0.85\columnwidth]{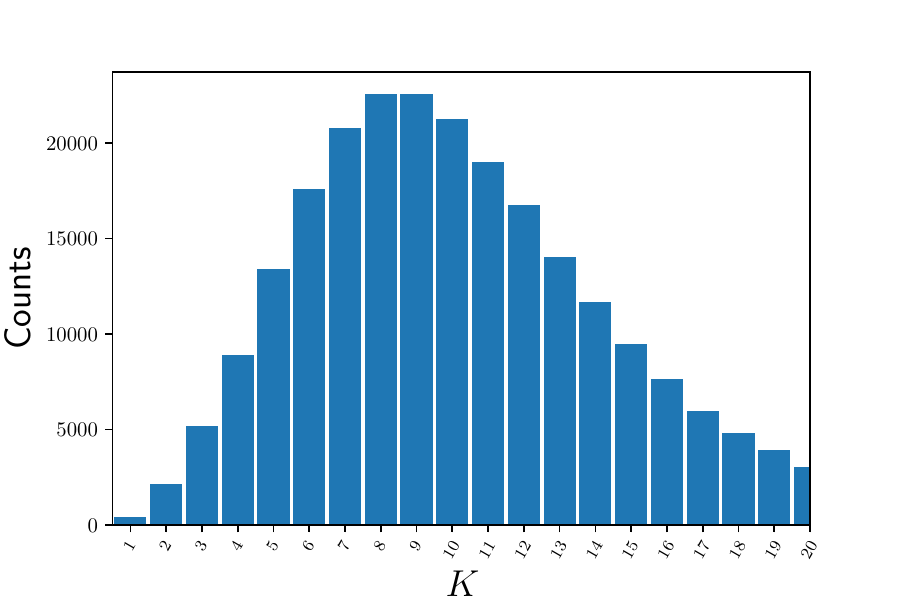}%
	\label{fig:new_K_histo}
}
\caption{Histograms of depths selected by {\sf mpNet} when equipped with $\mathtt{SC_1}$ and $\mathtt{SC_2}$.}
\label{fig:K_histo}
\end{figure}
  

\section{Experimental validation}
\label{sec:experiments}
\new{Let us now assess {\sf mpNet} on realistic synthetic channels. To do so, we consider the statistical spatial channel model (SSCM) channel model \cite{Samimi2016} in order to generate non-line-of-sight (NLoS) sparse channels at $28\,\text{GHz}$ (see \cite[table~IV]{Samimi2016}) corresponding to all users. This amounts to generating:
 \begin{enumerate}
	 \item the total number of paths $P$,
	 \item the DoAs $\{\overrightarrow{u_1},\dots,\overrightarrow{u_P}\}$ associated with each path ,
	 \item the complexe gains $\{\overrightarrow{\beta_1},\dots,\overrightarrow{\beta_P}\}$ associated with each path.
 \end{enumerate}
 Once generated, those parameters are used to compute the actual channels using equation \eqref{eq:multipath_channel}. We consider the same setting as in section~\ref{sec:impact}, namely a BS equipped with a ULA of $64$ antennas, with an half-wavelength nominal spacing and unit nominal gains used to build the imperfect nominal dictionary $\tilde{\mathbf{E}}$ (with $A=8N$) which serves as an initialization for {\sf mpNet}. The actual antenna arrays are generated the same way as in section~\ref{sec:impact}, using \eqref{eq:imperfection}. The generated physical parameters are kept fixed for the whole duration of each experiment, except for section \ref{xsec:anomaly} where we describe how they evolve.

 An important consideration is that the way we generate channels corresponds to a setting in which each channel observation is associated with a different user, simulating the initial access of users to the BS. In this way, our model does not need to be aware to which user the channel it is estimating corresponds. This is possible because the dictionary we consider at the BS is common to all users and each of its columns corresponds to the channel response to an excitation in a particular spatial direction (which is the same irrespective of which user is considered). This is in contrast with bayesian channel estimators such as the LMMSE, which require several channel observations of the same user in order to estimate its individual distribution to be injected as prior knowledge. }

\subsection{Fixed SNR}
\label{xsec:fixed_SNR} 
 
For the first set of experiments, we consider two model imperfections: $\sigma_p = 0.05,\,\sigma_g = 0.15$ (small uncertainty) and $\sigma_p = 0.1,\,\sigma_g = 0.3$ (large uncertainty) to build the unknown ideal dictionary $\mathbf{E}$. The input SNR is fixed during the experiment and takes the values $\{5,10\}\,\text{dB}$.

We compare the performance of various configurations of {\sf mpNet}, namely the version with a fixed number of iterations $K$, with $K$ set to $6$ for the SNR of $5\,\text{dB}$ and to $8$ for the SNR of $10\,\text{dB}$ (determined by cross validation) and the versions with an adaptive stopping criterion, using criteria $\mathtt{SC_1}$ and $\mathtt{SC_2}$ described in section~\ref{ssec:stopping_criterion}.
 In addition, the proposed method is compared to the LS estimator and to the OMP algorithm using the stopping criterion $\mathtt{SC_2}$, with either the imperfect nominal dictionary $\tilde{\mathbf{E}}$ or the unknown ideal dictionary $\mathbf{E}$. Finally, in order to demonstrate the advantages of using the imperfect model initialization, we compare it to using the well-known Xavier random initialization \cite{Glorot2010} which assumes no prior knowledge. \new{This baseline corresponds to a classical online dictionary learning method \cite{Mairal2010}.}
 
 \new{Note that a quantitative comparison with the MMSE estimator is absent from our experiments as it is not directly possible. Ideally, MMSE requires exact knowledge of the channel correlation matrix to apply. In the absence of that knowledge, an estimation procedure should be conducted which requires a very large number of time coherent samples to be satisfying. On the other hand, our method does not require prior information on the channel statistics since the prior information is entirely based on physics, which is an advantage compared to MMSE.}

\noindent{\bf Results.} The results of this experiment are shown on figure~\ref{fig:multi} as a function of the number of channels of the form~\eqref{eq:observations} seen by the BS over time. The performance measure is the normalized mean squared error ($\text{NMSE} = \Vert \hat{\mathbf{h}} - \mathbf{h} \Vert_2^2/\Vert \mathbf{h} \Vert_2^2$) averaged over minibatches of $200$ channels. 
Several comments are in order:
\begin{enumerate}[leftmargin=*]
	\item The imperfect model is shown to be well corrected by {\sf mpNet}s, the green and the two blue curves being very close to the red one (OMP with ideal unknown dictionary) after a certain amount of time. This is true both for a small uncertainty and for a large one and at all tested SNRs. Moreover, this shows that {\sf mpNet}, although based on MP, is capable of attaining a performance level close to that of OMP once it converges. Note that using the nominal dictionary (initialization of {\sf mpNet}) may be even worse than the LS method, showing the interest of correcting the model, since {\sf mpNet} always ends up outperforming the LS, thanks to the learning process.
	\item Comparing figures \ref{fig:multi_a} and \ref{fig:multi_b}, it is interesting to notice that learning is faster and the attained performance is better with a large SNR (the green and the two blue curves get closer to the red one faster), which can be explained by the better quality of data used to train the model.
	\item Comparing the figures \ref{fig:multi_a} and \ref{fig:multi_c}, it is apparent that a smaller uncertainty, which means a better initialization since the nominal dictionary is closer to the ideal unknown dictionary, leads to a faster convergence, but obviously also to a smaller improvement.
	\item Looking at the purple curve on all figures, it is apparent that initialization matters. Indeed, the randomly initialized network takes longer to converge and performs much worse than the one initialized with the nominal dictionary.
	\item The green and the two blue curves are all close to each other. In terms of performance, the light blue curve (corresponding to the use of $\mathtt{SC_2}$ as a stopping criterion) always leads to the best performance, followed by the green curve (corresponding to a fixed depth) and finally the dark blue curve (corresponding to the use of $\mathtt{SC_1}$). Figure \ref{fig:multi_b} shows that with a lower SNR, the gap between the dark blue and the rest of the {\sf mpNet} curves is more pronounced. To further understand the difference between $\mathtt{SC_1}$ and $\mathtt{SC_2}$, we show on figure~\ref{fig:K_histo} histograms of the selected depths for the experimental settings corresponding to the leftmost figure in figure~\ref{fig:multi}. We see that, for $\mathtt{SC_1}$, the distribution is centered at $K=5$ which is 3 iterations behind the network with the best results at a fixed depth. However, for $\mathtt{SC_2}$, the distribution is centered at $K=8$ which is the optimal value for a fixed depth. Moreover, the $\mathtt{SC_2}$ version outperforming the fixed depth one is a result of its adaptability to each channel observation. 
\end{enumerate}
These conclusions are very promising and highlight the applicability of the proposed method.
\balance

\subsection{Varying SNR}
\begin{figure}[t]
	\includegraphics[width=\columnwidth]{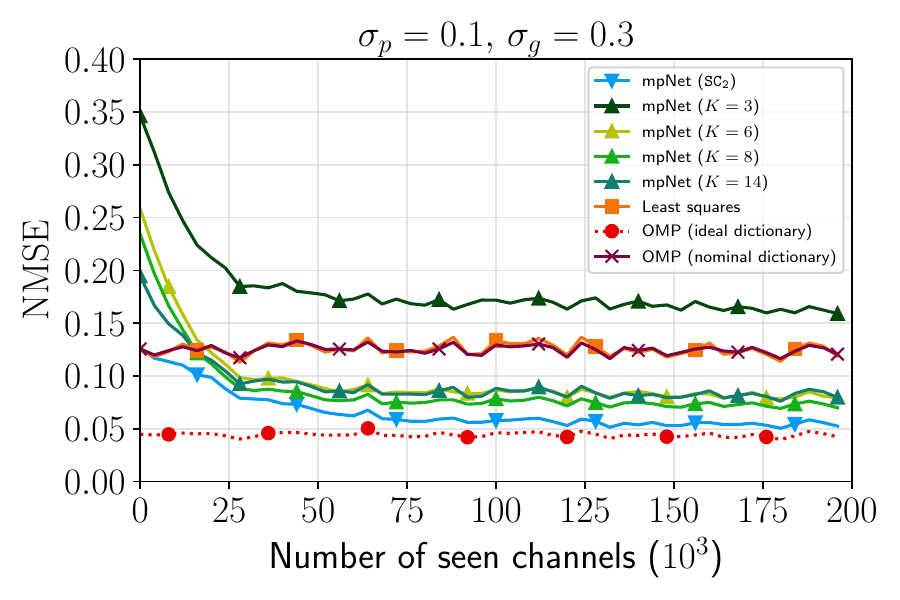}
	\caption{Channel estimation performance on synthetic realistic channels for a varying SNR.}
	\label{fig:variable_snr}
\end{figure}

In practical scenarios, the SNR is not fixed, and rather depends on the distance between the transmitter and the receiver as well as propagation conditions. This variability has to be taken into account for the learning algorithm to be efficiently optimized. Using $\mathtt{SC_2}$, which takes into account both the noise level and the signal intensity, suggests that our model is capable of automatically adapting to the SNR. To verify this claim, we consider the same SSCM channel model described in the previous section, but this time, the distance between the transmitter and the receiver is sampled from a uniform distribution ranging from $60\,\text{m}$ to $200\,\text{m}$. Our model is then compared to its fixed depth versions for a selected set of values of $K$ ($\{3,6,8,14\}$), as well as to the other previously presented estimation methods. The results are shown in figure~\ref{fig:variable_snr}. While all versions of {\sf mpNet} are capable of learning over time, the adaptive one is clearly outperforming the others throughout the whole experiment, starting once again at the same error level as OMP equipped with the nominal dictionary (the brown curve). Furthermore, compared to the precedent experiment at a fixed SNR, the gap between the fixed depth and the adaptive version is more pronounced. The performance of {\sf mpNet} with $\mathtt{SC_2}$ is indeed very close in terms of performance in this more realistic scenario to what is achievable with a perfect knowledge of the physical parameters (here represented by the red curve).

\begin{figure}[t]
\centering
	\includegraphics[width=\columnwidth]{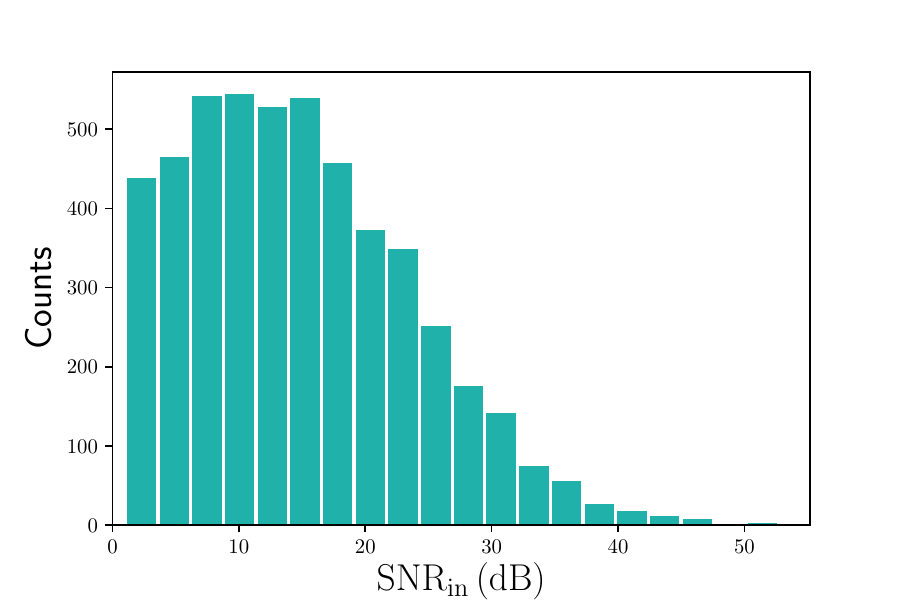}
	\caption{SNR distribution over 5000 generated channel observations.}
	\label{fig:snr_histo}
\end{figure}

Finally, figure~\ref{fig:snr_histo} shows how the SNR of the generated channel observations is distributed. Note that we choose not to consider observations with a low SNR ($<1\,\text{dB}$) as the model struggles to learn from very noisy instances. In practice, this could be achieved by setting a threshold on the intensity of received signals as a way to filter out observations heavily corrupted. This leads to a truncated normal distribution of the SNR centered at $10\,\text{dB}$.

\subsection{Anomaly detection and recovery}
\label{xsec:anomaly} 

\begin{figure*}[t]
	\captionsetup{font=footnotesize}
		\begin{subfigure}[b]{0.333\textwidth}
			\includegraphics[width=\columnwidth]{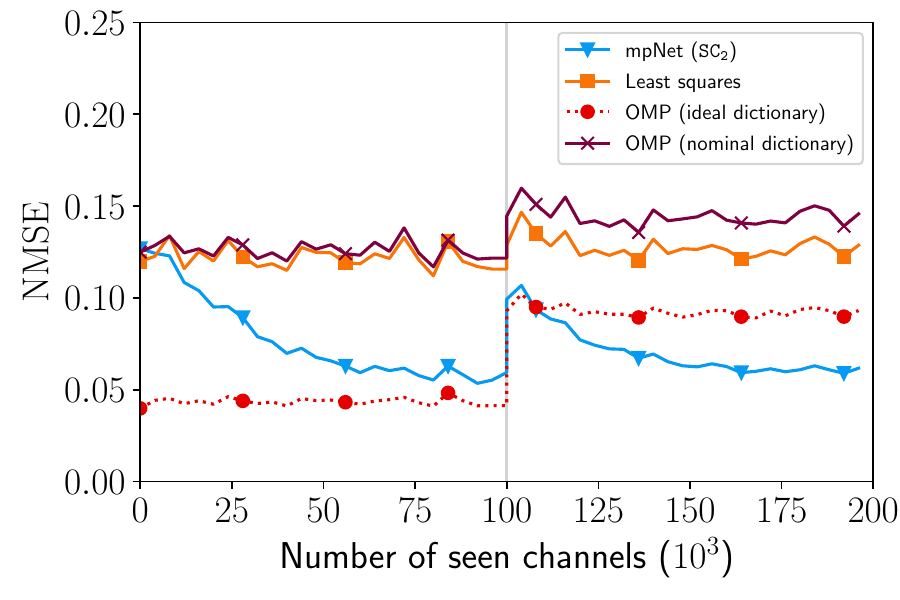}
			\caption{$\sigma_p={0.1},\, \sigma_g={0.3},\,\textrm{{broken antennas}}={10}\%$}
		\end{subfigure}
		\begin{subfigure}[b]{0.333\textwidth}
			\includegraphics[width=\columnwidth]{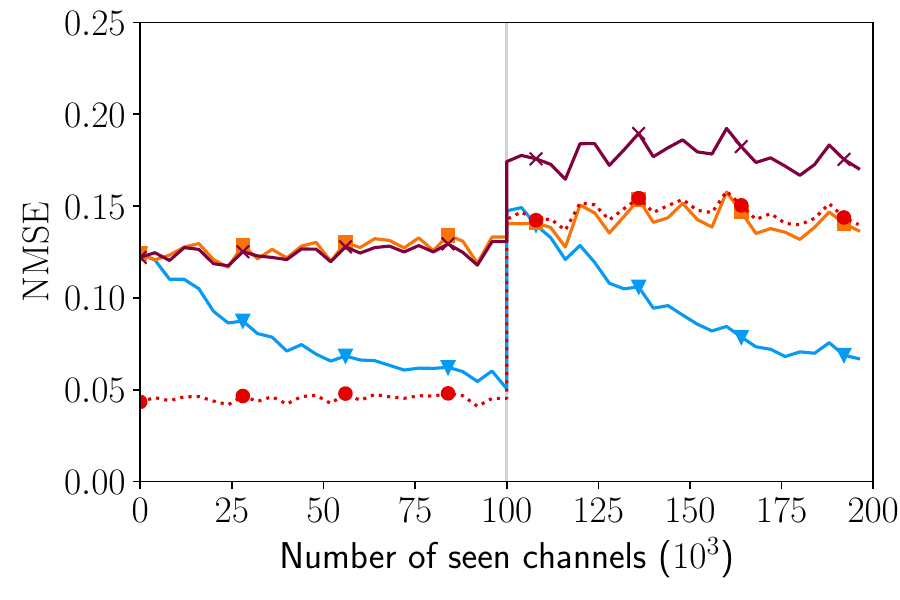}
			\caption{$\sigma_p={0.1},\, \sigma_g={0.3},\,\textrm{{broken antennas}}={30}\%$}
		\end{subfigure}
		\begin{subfigure}[b]{0.333\textwidth}
			\includegraphics[width=\columnwidth]{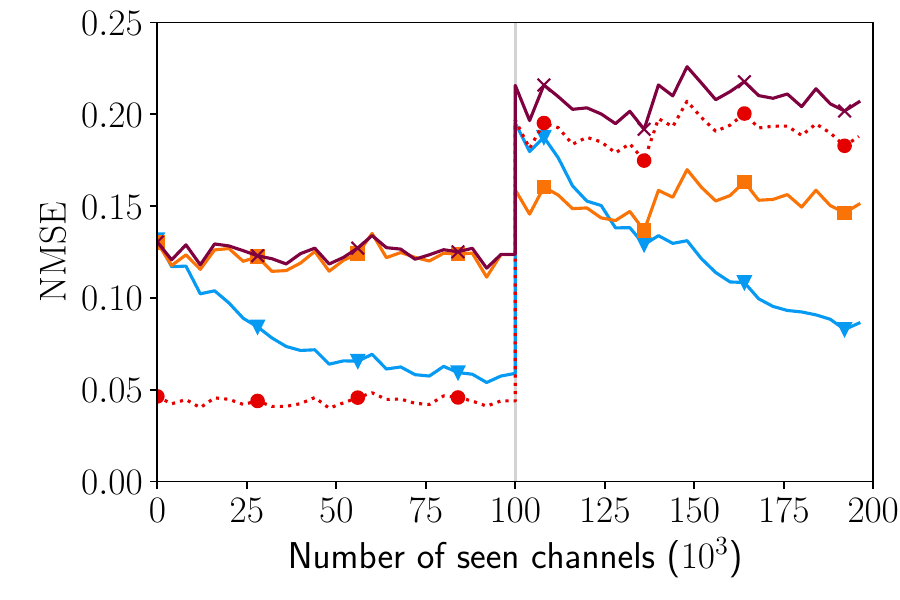}
			\caption{$\sigma_p={0.1},\, \sigma_g={0.3},\,\textrm{{broken antennas}}={50}\%$}
		\end{subfigure}
		\caption{Adaptation to antenna damage. The horizontal bar at the middle marks the moment at which the break happens.}
		\label{fig:break}
\end{figure*}

One of the benefits of online learning is the continuous adaptation of the model to incoming data. If this data is disturbed, the impact would be observed on the cost function used by the system. Indeed, the distribution of the new data would be different from the one on which the system has learned so far. The speed at which the error increases would be proportional to the rate at which the distribution of the new data shifts from its original state.  Thankfully, this increase in the error is simultaneously compensated by the training that is done on the network. This behavior may prove useful for detecting and recovering from anomalies. Anomalies could occur for many reasons in a massive MIMO system, including:
\begin{itemize}
	\item Bent or broken antennas due to natural causes
	\item Improperly adjusted antennas after an intervention
	\item Disoriented array due to wind 
\end{itemize}
In this section, we propose to test the ability of our model to detect and adapt to various types of anomalies. \new{Note that we assume that the antennas' parameters are only known to the BS at the beginning of its life and that any shift from the initial state cannot be detected directly.}

\noindent{\bf Out of order antennas.} Let us first simulate antennas that go out of order, in a BS equipped with 64 antennas. This is done by setting some of the antenna gains to zero at a certain point of time during training. \new{However, the BS will still receive noise on out of order antennas.} We consider 3 scenarios: 10\%, 30\% and 50\% of broken antennas (chosen uniformly at random). Similarly to previous experiments, the channels are generated following the SSCM model at a variable SNR (figure~\ref{fig:snr_histo}) and the adaptive version of {\sf mpNet} with $\mathtt{SC_2}$ is compared to the other estimation methods. Figure~\ref{fig:break} shows the results. It can be seen that the training starts as usual with our model slowly approaching the performance of OMP with the optimal dictionary. The anomaly can be observed at the middle of training when the number of seen channels reaches ${\sim} 100000$ channels. All the estimation methods see their error jump, with the exception of LS where no change is observed. The amount by which the error increases is proportional to the number of broken antennas. The error starts decreasing again for {\sf mpNet} as the training resumes, but naturally stays the same for OMP based methods that do not correct the dictionary they use. By the end of the experiment, and depending on the number of broken antennas, {\sf mpNet} can completely recover from the damage and its error reaches once again the level it successfully attained right before the anomaly.

The LS estimation method does not depend on any physical parameter, which explains the stable error level it maintained throughout the experiment. OMP methods, however, see their error increase because the physical parameters they are based on become less precise and thus induce a bigger error. Those methods are hence capable of detecting the exact moment where the damage happens but are incapable of adapting. On the other hand, and for the same reason, {\sf mpNet} is also capable of detecting the anomaly but rapidly adapts its parameters (dictionary). In practice, the detection could be implemented via a simple threshold.

Note that the anomaly recovery only concerns the channel estimation performance. Indeed, a BS with broken antennas will always be less efficient than a fully functional one, especially in terms of channel capacity. In summary, the model does the best it can on the channel estimation task with the available means.

\begin{figure*}[htbp]
	\captionsetup{font=footnotesize}
	\begin{subfigure}[b]{0.333\textwidth}
		\includegraphics[width=\columnwidth]{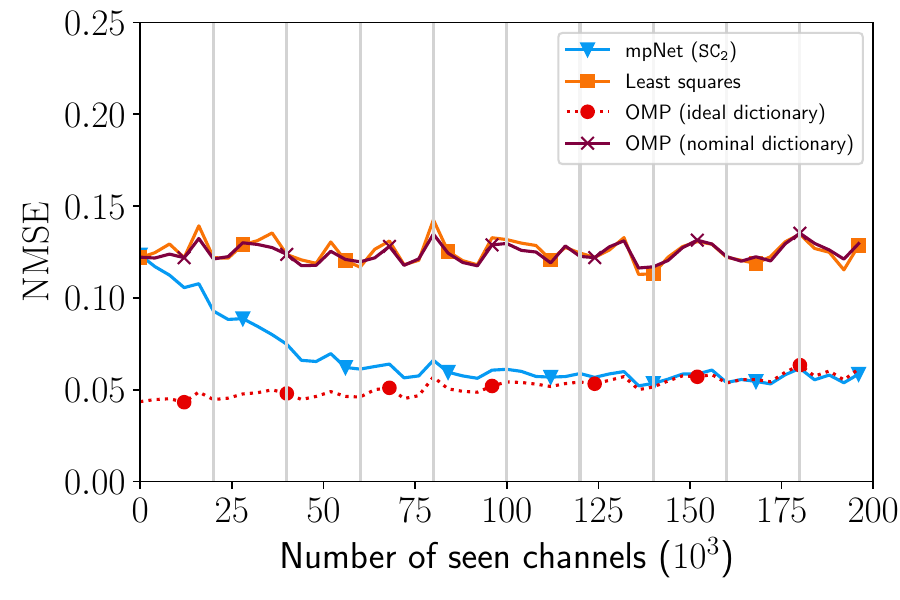}
		\caption{$\sigma_p={0.1},\, \sigma_g={0.3},\, \sigma_a={0.05}$}
	\end{subfigure}
	\begin{subfigure}[b]{0.333\textwidth}
		\includegraphics[width=\columnwidth]{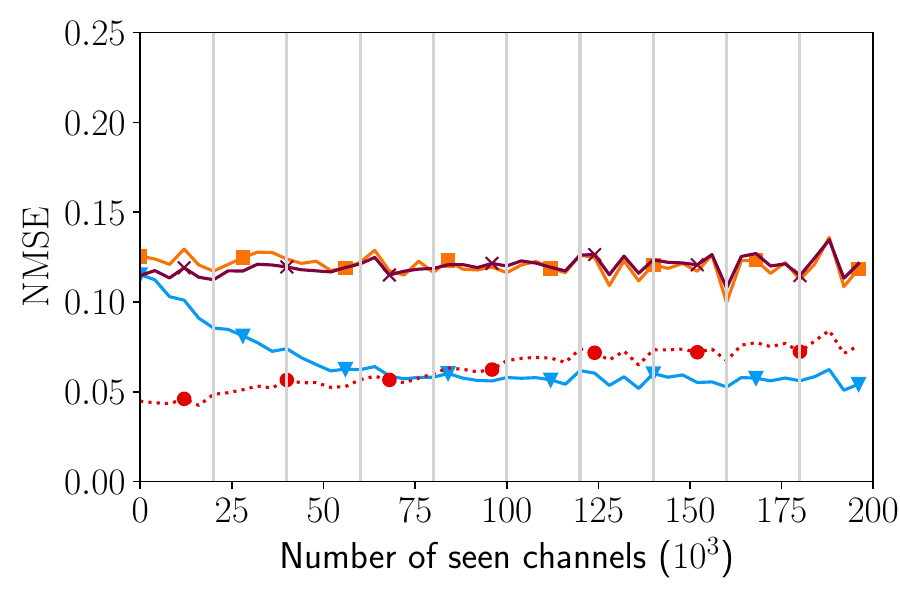}
		\caption{$\sigma_p={0.1},\, \sigma_g={0.3},\, \sigma_a={0.1}$}
	\end{subfigure}
	\begin{subfigure}[b]{0.333\textwidth}
		\includegraphics[width=\columnwidth]{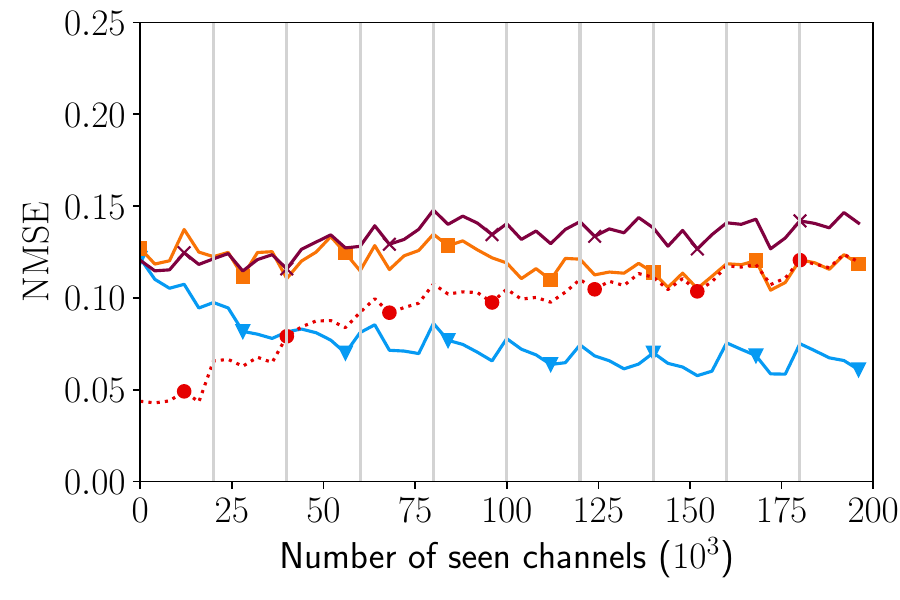}
		\caption{$\sigma_p={0.1},\, \sigma_g={0.3},\, \sigma_a={0.2}$}
	\end{subfigure}
	\caption{Adaptation to antenna aging. The horizontal bars mark the different moments at which the aging happens.}
	\label{fig:aging}
\end{figure*}

\noindent{\bf Aging antenna array.} The second type of anomaly we consider is antenna aging. \new{It corresponds to antenna gains slowly shifting away from their initial values, it could be due for example to slow change of temperature throughout the seasons, or change in weather conditions more generally.}

Considering antenna aging, using the initially ideal dictionary will lead to an increase of the error over time since the physical parameters are less and less precisely known. To simulate this phenomenon, we consider the same settings as for the precedent experiment. We then iteratively add noise to the antenna gains over the course of training. This is done for 10 iterations. Antenna gains at iteration $t$ are thus expressed as
\begin{equation}
g_{i,t} = g_{i,t-1} + n_{i}, \, n_{i} \sim \mathcal{CN}(0,\sigma_a^2),
\end{equation}
where $\sigma_a$ could be seen as a measure of the severity (speed) of aging. We consider 3 levels of aging: $\sigma_a=0.05$ (mild), $\sigma_a=0.1$ (medium) and $\sigma_a=0.2$ (severe). Again, channels are generated following the SSCM model at a variable SNR (figure~\ref{fig:snr_histo}) and the adaptive version of {\sf mpNet} with $\mathtt{SC_2}$ is compared to the other estimation methods. Results are shown on figure~\ref{fig:aging}. We observe that both OMP based estimations see their error progressively increase at a rate proportional to the severity of aging, but performance worsens more rapidly when starting with an ideal dictionary. On the other hand, the neural network continues to learn and the impact of aging is barely noticeable.  {\sf mpNet} online learning compensates for the error induced by the continuous change of physical parameters. Finally, no noticeable change is observed on the LS estimation method.

\subsection{From ULA to UPA}

\begin{figure}[htbp]
	\includegraphics[width=\columnwidth]{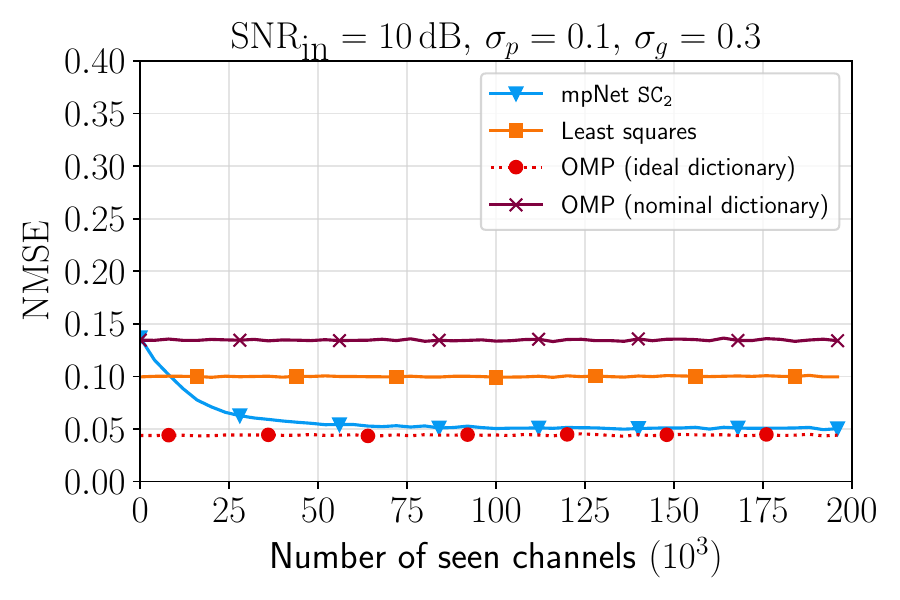}
	\caption{Channel estimation performance on synthetic realistic channels for a BS equipped with a UPA.}
	\label{fig:UPA}
\end{figure}

The physical model on which is based {\sf mpNet} is structure-agnostic, meaning that it is meant to work with any antenna array structure. This suggests that our model, which was initially tested on ULAs, is capable of working with any structure as well. To verify this claim, we propose to adapt it to uniform planar arrays (UPAs). A change in the way steering vectors are generated is required to take into account the rotational symmetry that was verified for ULAs and that no longer holds for UPAs. Therefore, the dictionary of steering vectors has to be built from DoAs sampled from the whole 3D half space, instead of a half-plane in the case of ULAs.

We conducted an experiment similar to the ones described in section~\ref{sec:experiments}. We consider an UPA consisting of a square grid of $8\times8$ antennas ($N=64$) separated by half-wave lengths and placed on the $xz$-plane . Channels are generated at a fixed SNR of $10\,\text{dB}$. To take into account the additional dimension of UPAs, ideal antenna positions are this time given by
\begin{equation}
\overrightarrow{a_i} = \tilde{\overrightarrow{a_i}} + \lambda\mathbf{n}_{p,i}, \, \mathbf{n}_{p,i} = {\small\begin{pmatrix} e_{x,p,i}, &0, & e_{z,p,i} \end{pmatrix}^T},
\end{equation}
with $e_{x,p,i},e_{z,p,i} \sim \mathcal{N}(0,\sigma_p^2)$.
We compare the adaptive version of {\sf mpNet} equipped with $\mathtt{SC_2}$ to other estimation methods. The results are shown on figure~\ref{fig:UPA}. Once again, the model successfully learns over time reducing its estimation error while the other classical methods maintain their performances.

This small experiment can be seen as a sanity check to show that, indeed, the model is capable of accepting any antenna structure with minimal changes in the implementation.


\section{Conclusion and perspectives}
\label{sec:conclusion}

In this paper, we introduced {\sf mpNet}: a neural network allowing adding flexibility to physical models used for MIMO channel estimation. It is based on the deep unfolding strategy that views a classical algorithm (matching pursuit in this case) as a neural network, whose parameters can be trained. The proposed method was shown to correct incrementally (via online learning) an imperfect or imperfectly known physical model in order to make channel estimation as efficient as if the unknown ideal model were known. It is trained in an unsupervised manner as an autoencoder, and {\sf mpNet} can be seen as a denoiser for channel observations. Training {\sf mpNet} thus does not necessitate a database of clean channels, nor an offline training phase, which makes it particularly attractive for practical systems and unalike previously proposed methods.


We have shown that initializing the network with a dictionary of imperfect steering vectors (as opposed to using a random initialization) improves performance considerably. In the experimental part of the paper, we simulated the model imperfection by introducing uncertainties on the antenna gains and positions, but it is important to highlight the fact that the method could in principle correct many other model imperfections (such as uncertainties about the antenna diagrams, couplings between neighboring antennas, etc.).

Moreover, we introduced a stopping criterion, inspired by previous work on the OMP algorithm, to dynamically select the optimal depth of {\sf mpNet}. This was shown to be particularly convenient when working on observations with a varying SNR level, which more accurately resembles real world channel observations. Evaluated on realistic synthetic data, this approach showed great results compared to other methods.

In addition, online learning enabled us to exploit the observed change in data distribution following an anomaly occurrence to detect and recover from it. We simulated two types of anomalies: antenna damage and aging. In both cases, our model was capable of efficiently recovering from the decrease in performance over time.

Finally, we proved that our model is capable of adapting to any antenna structure with no apparent drop in performance or change in behavior. In particular, we showed that a simple change in the steering vector generation process was required to adapt the model, previously based on a ULA structure, to a UPA structure.

\new{In future work, we could explore the unfolding of more sophisticated sparse recovery algorithms (such as iterative soft thresholding \cite{Daubechies2004} or AMP \cite{Donoho2009}) using the same strategy in a way that would mitigate their relatively high computational complexity, making them suitable for online training.} In addition, other stopping criteria could be integrated to the model and tested. \new{Also, we could drop the isotropic antennas assumption and consider anisotropic antennas where complex gains would depend on the DoA, in which case signals arriving at DoAs where gain is stronger are to be prioritized. The proposed method could also be extended quite readily to handle multi-antenna users by structuring the dictionary used within {\sf mpNet} as a Kronecker product of steering vectors dictionaries. Another avenue worth exploring is the use of our model together with MMSE estimation. One could imagine using {\sf mpNet} to obtain cleaner channels than LS estimates when a user first connects to the BS in order to form an empirical covariance matrix that could be used within the framework of MMSE estimation later.} Finally, \new{it is worth noting that} the deep unfolding of the matching pursuit algorithm initialized with a dictionary based on an imperfect model is by no means limited to the MIMO channel estimation task and could be exploited for other tasks, as long as an initial model is available.

\bibliographystyle{unsrt}
\bibliography{biblio}

\end{document}